\documentclass[conference]{IEEEtran}
\IEEEoverridecommandlockouts
\ifCLASSINFOpdf
  \usepackage[pdftex]{graphicx}
  \graphicspath{{./figures/}}
  \DeclareGraphicsExtensions{.pdf,.jpg,.png,.xps}
\else
  \usepackage[dvips]{graphicx}
  \graphicspath{{./}}
  \DeclareGraphicsExtensions{.eps}
\fi
\usepackage{cite}
\usepackage{amsmath,amssymb,amsfonts}
\usepackage{algorithmic}
\usepackage{textcomp}
\usepackage{xcolor}
\usepackage{subcaption}
\usepackage{multirow}
\usepackage{soul}
\def\BibTeX{{\rm B\kern-.05em{\sc i\kern-.025em b}\kern-.08em
    T\kern-.1667em\lower.7ex\hbox{E}\kern-.125emX}}

\begin{document}

\title{Engineering a QoS Provider Mechanism for Edge Computing with Deep Reinforcement Learning
\thanks{This work has been partially performed in the framework of mF2C project funded by the European Union's H2020 research and innovation programme under grant agreement 730929.}
}

\author{
\IEEEauthorblockN{Francisco Carpio \textsuperscript{1}, Admela Jukan \textsuperscript{1}, Rom\'an Sosa \textsuperscript{2} and Ana Juan Ferrer \textsuperscript{2}}
\IEEEauthorblockA{\textsuperscript{1} Technische Universit{\"a}t Braunschweig, Germany, Email:\{f.carpio, a.jukan\}@tu-bs.de}
\IEEEauthorblockA{\textsuperscript{2} ATOS Research and Innovation, Spain, Email:\{roman.sosa, ana.juanf\}@atos.net}
}

\maketitle

\begin{abstract}
 With the development of new system solutions that integrate traditional cloud
 computing with the edge/fog computing paradigm, dynamic optimization of service
 execution has become a challenge due to the edge computing resources being more
 distributed and dynamic. How to optimize the execution to provide Quality of
 Service (QoS) in edge computing depends on both the system architecture and the
 resource allocation algorithms in place. We design and develop a QoS provider
 mechanism, as an integral component of a fog-to-cloud system, to work in
 dynamic scenarios by using deep reinforcement learning. We choose reinforcement
 learning since it is particularly well suited for solving problems in dynamic
 and adaptive environments where the decision process needs to be frequently
 updated. We specifically use a Deep Q-learning algorithm that optimizes QoS by
 identifying and blocking devices that potentially cause service disruption due
 to dynamicity. We compare the reinforcement learning based solution with
 state-of-the-art heuristics that use telemetry data, and analyze pros and cons. 
 
\end{abstract}


\section{Introduction}

% scenario
New emerging edge-based computing systems, also referred to as fog computing,
are designed to provide cloud computing capabilities closer to the users, in
order to reduce latency and network traffic by processing and storing data
locally. Whereas in cloud computing resources are centralized and static, in
edge computing, the heterogeneity and dynamicity of edge devices make the
orchestration of services an open challenge. QoS provisioning in edge computing
not only needs to address the dynamicity of resources but it also needs to deal
with the service disruptions and a variety of different hardware solutions for
service execution. In this new scenario, recent efforts have focused on
architecture and new algorithms, including learning based methods, to address
the challenges of resource allocation and QoS guarantees.

% problem
Quality of service is a known challenge in cloud computing due to hardware
failures (servers, links, switches) or software reconfigurations (e.g. Virtual
Machine (VM) migrations), for which mechanisms exist to maintain  a certain
level of QoS \cite{Faniyi2015}. In edge computing, on the other hand, these
mechanisms cannot be directly applied not only because the failures occur more
often and at different time scales, but also because of the dynamicity of the
connectivity between resources and difficulties in providing back up resources
dynamically. In these scenarios, the distributed nature of service execution
makes telemetry based solution a challenge, and while machine learning is a
valid option, and the open question is which machine learning solutions are
better suitable to consider the intrinsic dynamicity of resources in edge
computing systems.

% solution
We engineer a Deep Reinforcement Learning (DRL) based solution to optimize QoS
provisioning and present study in this paper of the measurements of quality of
this solution in dynamic edge computing networks. Specifically, we develop a
Deep Q-learning algorithm based on deep neural networks that is able to block or
allow the usage of devices for executing services in order to avoid SLA
violations in case of devices fail during the runtime. We choose reinforcement
learning since is particularly well suited for dynamic environments where the
algorithm has to adapt the decision process over time without requiring
pre-training sessions. Our algorithm has been designed to work as an integral
component, called \emph{QoS provider}, of an open source fog-to-cloud management
system developed under our ongoing mF2C project \cite{mf2c}. We compare our
solution with a heuristic algorithm that blocks devices based on availability
probabilities based on telemetry in the system and study pros and cons. 

The rest of the paper is organized as follows. Section II presents related work.
Section III describes the system architecture, while Section IV shows the deep
reinforcement learning approach. Section V analyzes the performance and section
VI concludes the paper. 

\section{Related Work}

% edge/fog computing 
Recently, a few ongoing projects and standardization frameworks, such as  our
ongoing project \cite{mf2c} and \cite{openfog} have started to materialize edge
computing solution into open source developments. In edge (fog) scenarios, the
traditional QoS provisioning models as used in the cloud are not suitable, and
hence new solutions are being designed. For instance, \cite{Kan2018} and
\cite{Song2017} propose task offloading methods to fulfill QoS requirements in
distributed edge computing. Similarly, \cite{Huang2017} proposes a QoS
provisioning mechanism for fog computing that is able to dynamically define
fine-grained QoS policies. No ongoing work however  has used reinforcement
learning for QoS provisioning. On the other hand, the idea of using
reinforcement learning for QoS control \emph{per se} in distributed systems is
not generally not new \cite{Li2010} and further research is needed to adapt the
previous finding to edge computing. 
 
% reinforcement learning
It should be noted that recently reinforcement learning has started to permeate
the areas of edge computing other than QoS. One of these areas is proposed in
\cite{Khelifi2019}, to solving the server and network resource allocation
problem \cite{Wang2019}. Related to QoS, paper \cite{Abundo2015} studies the
bidding decision process that an application provider would perform to ensure a
minimum throughput to guarantee QoS, by modelling the problem as a Q-Learning
problem constrained by multiple input parameters. \cite{Lin2016} proposes a
QoS-aware adaptive routing algorithm for distributed multi-layer control plane
SDN architectures. In cloud computing, the authors in \cite{Wei2018} have proven
the usefulness of reinforcement learning for building intelligent QoS-aware job
schedulers. However, to the best of our knowledge, ours is the first attempt to
engineer a QoS provisioning in edge computing networks with reinforcement
learning. By learning in runtime doing trial and error, we expect to improve QoS
in the scenarios where traditional telemetry based heuristics do not perform
well.  

\begin{figure}[!t]
  \centering
  \includegraphics[width=1.0\columnwidth]{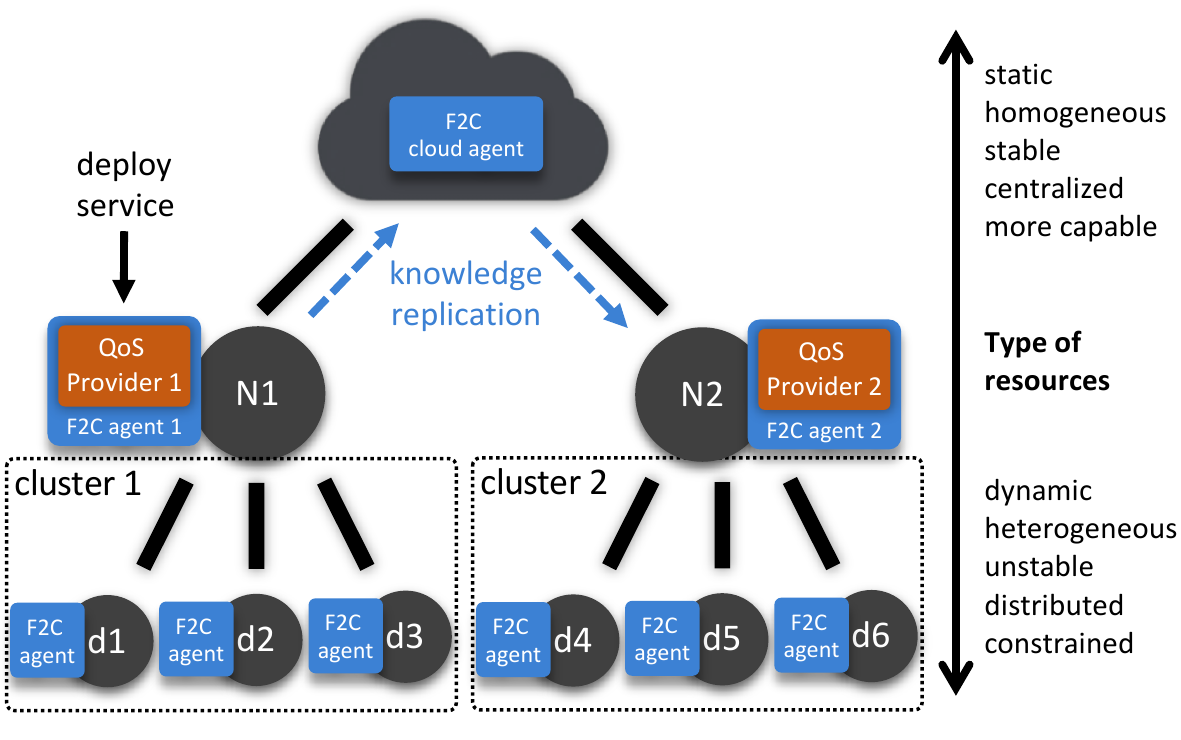}
  \caption{Edge/Fog to Cloud architecture}
  \vspace{-0.3cm}
  \label{arch}
\end{figure}

\section{System architecture}

This section first describes the reference architecture based on a typical
fog-to-cloud management system. We then
deep dive into the specific modules involved on QoS provisioning, including
\emph{Service Manager}, \emph{SLA manager} and \emph{QoS Provider}.

\subsection{Fog-to-Cloud system architecture}

\begin{figure}[!t]
  \centering
  \includegraphics[width=1.0\columnwidth]{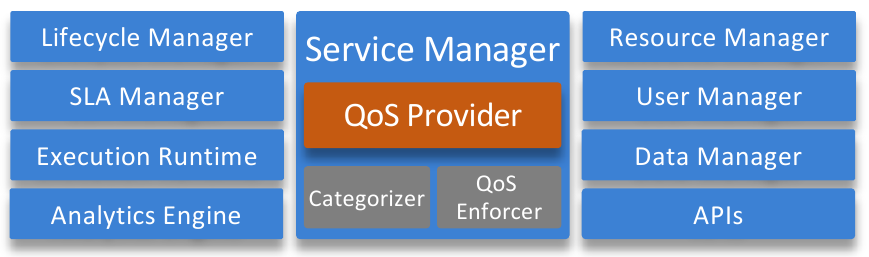}
  \caption{An F2C agent} 
  \vspace{-0.6cm}
  \label{system-arch}
\end{figure}

The design and development of the \emph{QoS Provider} has been performed as part
of the fog to cloud architectural platform developed in \cite{mf2c}, which by
itself is heavily leaning on the fog-to-cloud architectures proposed in Open Fog
Consortium standardization body. This architecture considers a hierarchical tree
topology of the overall system, where computing devices are connected at
different layers according to their compute and storage capabilities and their
connectivity. The more static and computationally capable devices are clustered
closer to the cloud, while the more dynamic and constrained devices clustered
are closer to the bottom of the hierarchy (fog). A simplified representation of
this architecture is shown in Fig. \ref{arch}. In this architecture, every
device (or logical cluster) runs a mF2C agent, where depending on the layer in
the hierarchy, the agent is expected to play different roles. For instance, in
the example with three layers shown in Fig. \ref{arch}, the intermediate nodes
\emph{N1} and \emph{N2} act as leaders of cluster 1 and 2, respectively. Here,
if a service is to be deployed in \emph{N1}, it will use the resources in
cluster 1, and \emph{QoS Provider 1} is responsible to provide certain QoS to
that service. Although the clustered devices \emph{d1}, \emph{d2} and \emph{d3}
also have the same F2C agent deployed, QoS provider does play any role there
since they are not leaders of the cluster. On the other hand, the cloud agent
here only acts as a backup entity for the acquired knowledge in \emph{F2C agent
1} to be replicated into other nodes. The generic architecture of every agent is
implemented with various functional modules, as shown in Fig. \ref{system-arch}.
Before starting the description of the \emph{QoS provider} as part of
\emph{Service Manager}, let us first explain how \emph{SLA manager} works in the
architecture, which is relevant to QoS (other modules are out of scope and can
be found in \cite{mf2c}). 

The SLA Manager implements the creation, storage and evaluation of the
agreements. An SLA agreement is a document that declares the QoS guarantees that
a service provider offers to a client; as such, the document contains the
involved parties, a description of the provided service and a set of guarantee
terms. The schema of the mF2C SLA agreements is based on the WS-Agreement
specification \cite{Andrieux2004}, using a JSON format. This facilitates the
management of SLAs by the devices with low computing capabilities that may be
present in a clustered area. In particular, the management of SLAs is done by
the leader of a cluster area, which is considered the device with higher
computing resources. Ideally, each device executes the SLA Management, so it can
take over the role of SLA manager in case the leader becomes unaccessible.The
guarantee terms in an agreement define the Service Level Objectives (SLOs) that
the service provider must fulfill. They are expressed as a constraint on a QoS
metric (e.g., service availability of 99.999\%). Our architecture considers
metrics at the level of the application and the infrastructure. An example of
QoS metric at the level of infrastructure is the availability of the devices,
while an example at the level of the application is the response time to execute
a given operation.  For the evaluation, the SLA Management relies on the
monitoring metrics provided by the Telemetry and the Distributed Execution
Runtime components. The actual value for the metric expressed in the SLO
constraint is compared to the threshold, raising an SLA violation when the
constraint is not satisfied. The guarantee term, besides the SLO, may define the
penalty that applies in case of a violation (e.g. a discount).

\begin{figure}[!t]
  \centering
  \small
  \algsetup{indent=2em}
  \begin{algorithmic}[1]
  \STATE N, f = 0
  \STATE B = N - (N $\cdot$ a)
  \STATE env = initializeEnvironment(N)
  \STATE availabilities = initialize(N)
  \LOOP
    \STATE env $\leftarrow$ blockDevices(env, availabilities, B)
    \STATE service = executeService(env)
    \STATE f = checkFailure(service)
    \IF{f = 1}
      \STATE failures = getFailedDevices(env)
      \STATE failProb = updateFailureProb(env, failures)
    \ENDIF
  \ENDLOOP
  \end{algorithmic}
  \caption{Telemetry based heuristic (TEL) algorithm}
  \vspace{-0.3cm}
  \label{heu_algo}
\end{figure}

\subsection{Service Manager with QoS Provider}

The \emph{QoS provider} component is part of a Service Manager module which is a
component software of mF2C \cite{mf2c}. Apart from the \emph{QoS provider}, the
\emph{Service Manager} is also composed by the \emph{Categorizer} and \emph{QoS
Enforcement}. The Categorizer is responsible of registering and categorizing new
services into the system, where a service is defined by different parameters
such as the application to run, the SLA, the minimum set of devices to run a
service, among others. The \emph{QoS Enforcement} is responsible to add new
devices for service execution in runtime in case the system predicts that are
not enough resources to fulfill the SLA agreement. When a service is executed
for the first time, the Service Manager generates a new QoS model for that
specific agent that is going to be executed in a set of specific devices. 

The \emph{QoS provider} module tries to assure that the SLA agreements are
fulfilled by blocking or allowing the usage devices based on their availability.
Because the \emph{QoS provider} does not run in runtime (like the \emph{QoS
enforcer}), this decision has to be made in advance, before the execution of a
service. For taking this decision, the \emph{QoS provider} makes use of
telemetry data that determines which agents failed in past executions and tries
to avoid their usage. The pseudo code of the telemetry based heuristic algorithm
(TEL) is shown in Fig. \ref{heu_algo}. $N$ is the total number of devices and
\emph{B} is the number of devices to block in every iteration. \emph{B} is
determined based on \emph{N} and \emph{a} (acceptance ratio). This acceptance
ratio is the minimum percentage of required devices for a specific service to
run properly. We initialize the environment, which consists of an array of
booleans ($d_1$, $d_2$, ..., $d_n$) each one representing a device in a cluster
where an instance of the algorithm is running, and the availabilities by
specifying the number of agents. Then, we enter in the loop that is run for
every service execution. Inside the loop, we call a function to block devices
specifying the environment, the availabilities and the number of devices to
block. After the service is executed specifying which devices can be used, the
algorithm checks whether the service was disrupted or not from telemetry data
provided by \emph{Analytics Engine} (see Fig.\ref{system-arch}). In case of
disruption (\emph{f = 1}), we check which devices were not available during the
execution and we update the new availabilities of each device.

\begin{figure}[!t]
  \centering
  \includegraphics[width=0.8\columnwidth]{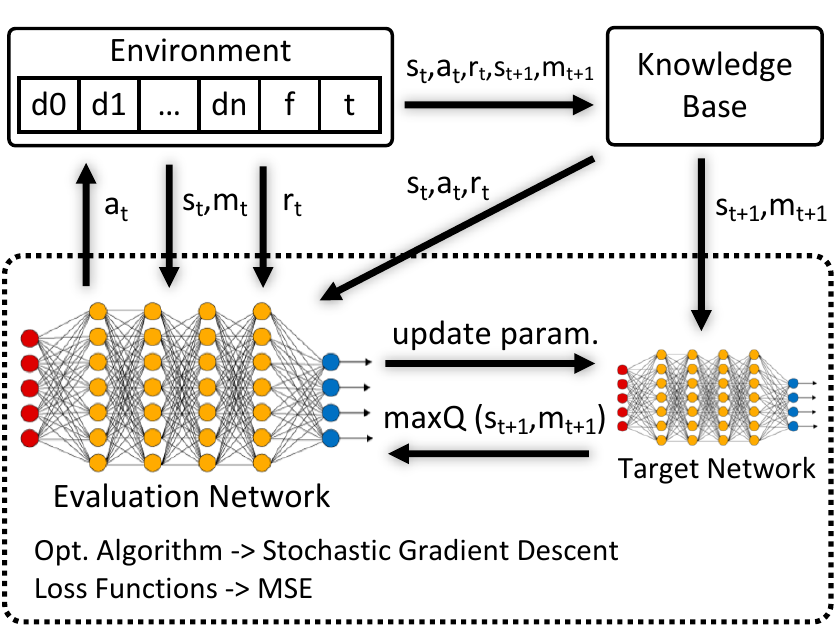}
  \caption{Deep Reinforcement Learning model}
  \vspace{-0.2cm}
  \label{model}
\end{figure}

\section{Integrating Deep Reinforcement Learning (DRL) into the QoS provider}

In reinforcement learning models, an agent takes some actions in an environment
based on observations, where later those actions are rewarded back to the agent.
The objective is to maximize the cumulative reward through performing actions,
in some cases prioritizing short term rewards or in other cases looking forward
to future long term rewards. Fig. \ref{model} represents our model design. The
environment consists of an array of booleans ($d_1$, $d_2$, ..., $d_n$), each
one representing a device in a cluster (see Fig. \ref{arch}), plus an specific
boolean $f$, used to indicate that the service execution failed, plus an integer
$t$ to determine the time step. Each boolean represents whether a specific
device $d_x$ is blocked for allocation (value 1) or allowed (value 0).
Therefore, considering $N$ the total number of devices, the length of the
environment is equal to $N + 2$. From this environment, which will be modified
into a new state on every iteration, a new input is generated for the evaluation
network. Because in every iteration $t$, a new state $s_t$ from the environment
is generated, a new reward $r_t$ is calculated based on the previous action and
the weights of the network updated. The number of outputs is determined based on
the number of devices to allow or block for allocation. Since the individual
decision of blocking and allowing a certain device are probabilistically
independent, two actions are possible for every device, plus the action of not
performing any action at all. Therefore, the output length of the model is equal
to $2N + 1$. From the output values, the algorithm selects one action $a_t$
based also on a mask $m_t$. The mask has the same length as the output and is
used to limit the actions the agent can take for an specific environment. The
next example, when $N=2$, shows the simplified procedure of the algorithm:
\begin{subequations}
  \begin{align}
   environment(s_{t}) & = [1,0,0,2] \\
   mask(s_{t}) & = [1,0,0,1,1] \\
   output(s_{t}) & = [2.11, 3.02, 1.55, 0.053, 0.12] \\
   action(s_{t}) & = 0 \\
   environment(s_{t+1}) & = [0,0,0,3]  \\
   reward(s_{t+1}) & = r
  \end{align}
\end{subequations}

In this example, the environment length is 4, where the two first positions
indicate whether devices 1 and 2, respectively, are allowed (0) or blocked (1),
the third position indicates whether the service is disrupted or not, and the
last position indicates the time step. Then, a mask at the state $s_t$ is
generated, setting array elements to 0 when an action cannot be performed; and
to 1, otherwise. In this specific example, considering the mask has $K = 5$
elements and based on the environment($s_t$), the element $k=0$ determines if
the first device from the environment can be switched to \emph{allowed}. Because
in the environment that device was \emph{blocked}, for this iteration we set the
element $k=0$ to 1 to indicate that the device can be allowed. However, because
it was already blocked, the element $k=1$ of the mask is set to 0, to not allow
the algorithm to block that device again. The same procedure applies for the
second device, where the elements $k=2$ is set to 0 and $k=3$ is set to 1, to
specify to the algorithm that it can block that device but not allowing it
again. This procedure is necessary to maintain each \emph{allow} and
\emph{block} probabilities independent for every device providing more knowledge
to the model for the decision process. In case the optimum decision is to not
perform any action on a specific state, the last element of the mask is used for
that purpose, being always set to 1. Then, based on the mask($s_t$), an action
is taken at instance $s_{t}$, according to the values from the output array. The
action is equal to the position of the array with the maximum value that the
mask allows to use. In this case, the maximum value from the output that the
mask allows to use is $2.11$, so the action is equal to that position in the
array, i.e. 0. This action modifies the environment for the next time step, by
switching from 1 to 0 the first element of the environment. It is to be noted
that the last element of the environment is just the time step counter; in this
case it indicates that is the 4th iteration of the algorithm. At that point a
reward($s_{t+1}$) is calculated taking into account the current state of the
environment and the model updates the network values for the next iteration.
 
The algorithm consists of a function $Q$ that calculates the quality of actions
in different combinations of states. So, at each state $s_t$, the agent chooses
an action $a_t$, observes a reward $r_t$ and updates into an new state
$s_{s+1}$. This process updates iteratively the function $Q$ following the next
equation:
\begin{equation}
  Q_{t+1}(s_t, a_t) = Q_t(s_t, a_t) + \alpha[R(s_t, a_t) + \gamma \cdot \max Q(s_{t+1}, a_{t+1})]
\end{equation}
, where $Q_{t+1}(s_t, a_t)$ is the updated value for next iteration and
$Q_t(s_t, a_t)$ is the old Q value. The $\alpha$ value is the learning rate ($0
< \alpha \leq 1$) which determines the weight between the new information and
the previous one. The closer $\alpha$ value is to 0, the less new information
the agent learns, while the closer to 1, the more new information the agent
leans. $R(s_t, a_t)$ is the reward observed after performing an action $a_t$ in
state $s_t$. $\gamma$ ($0 < \gamma \leq 1$) is the discount factor which
determines the importance of future rewards in comparison with immediate
rewards. A $\gamma$ close to 0 makes future rewards worth less than immediate
values, while a value close to 1 makes future rewards worth as much as immediate
rewards. The $\max Q(s_{t+1}, a_{t+1})$ value is the maximum estimated future
reward given the new state $s+1$ and the possible actions for that specific
state $a_{t+1}$. To be observed, that this value is calculated from the target
network which is a copy of the evaluation network, but which parameters are only
updated at certain frequency and not in every step as for the evaluation
network. This is done to improve convergence of training and stability to the
model. Both neural networks will be updated by stochastic gradient descent and
will use Mean Square Error as the loss function.

\begin{figure}[!t]
  \centering
  \small
  \algsetup{indent=2em}
  \begin{algorithmic}[1]
  \STATE N, action = -1, f = 0, t = 0, r = 0
  \STATE env = initializeEnvironment(N, f, t)
  \STATE nextEnv = initialize(N, f, t + 1)
  \STATE mask = generateMask(env)
  \LOOP
    \STATE service = executeService(env)
    \STATE nextEnv $\leftarrow$ f = checkFailure(service)
    \STATE reward = computeReward(nextEnv)
    \STATE nextMask = generateMask(nextEnv)
    \STATE addExp(env, nextEnv, action, reward, nextMask)
    \STATE trainNetwork()
    \STATE env = nextEnv
    \STATE mask = generateMask(env) 
    \STATE action = getAction(env, mask)
    \STATE nextEnv = modifyEnvironment(env, action)
  \ENDLOOP
  \end{algorithmic}
  \caption{Deep Reinforcement Learning (DRL) algorithm}
  \vspace{-0.3cm}
  \label{drl_algo}
\end{figure}

A pseudo algorithm is shown in Fig. \ref{drl_algo}. Before starting to iterate,
the environments, \emph{env} and \emph{nextEnv}, are initialized by specifying
the number of devices (N) in the cluster. Also, an initial \emph{mask} is
generated from the initial environment, following the procedure previously
mentioned. From this point, the algorithm enters in a loop where every iteration
is an execution of the specific service. For every iteration, the first step is
to check if the service failed in the previous iteration; if so, the
\emph{nextEnv} is modified by switching the $f$ value to 1. Then, a reward is
calculated based on \emph{nextEnv}, where the status of each device (blocked or
allowed) is checked, while considering if the service failed or not, and points
are given according to that. The total reward is the summation of points per
device having 4 different cases: 1) $+10$ points, if a device was allowed and
service did not fail, 2) $-10$ points, if a device was allowed and service
failed, 3) $0$ points, if a device was blocked and service did not fail and 4)
$-10$ points, if a device was blocked and the service failed. With this reward
function, we are positively rewarding the cases where more devices are allowed
and no service failures occur. On the contrary, we are penalizing the cases
where there is a service failure without differentiating whether a certain
device was allowed or blocked. Finally, we are not rewarding at all the cases
where devices are blocked and no service failure ocurred. While this last case
is positive, the objective is to maximize the number of used devices, therefore,
by not rewarding we are pushing the model to try to allow devices as long as
they do not cause service failures. Then, a new mask \emph{nextMask} based on
the \emph{nextEnv} is created and the model adds a new experience to the
knowledge base, where the \emph{env}, the \emph{nextEnv}, the action, the reward
and the \emph{nextMask} are specified (see also Fig. \ref{model}). The next step
is training the evaluation network which consists on calculating the function
$Q$, previously explained, per a batch of experiences, where the
\emph{batch\_size} is a parameter value. This training will only occur after the
number of experiences is greater than an initial value \emph{start\_size}. The
maximum number of experiences that the model can store in the knowledge base is
determined by a \emph{memory\_capacity} value. When a new experience is created
and the memory is full, a randomly old experience is removed to let the new
experience be added. This is done to limit the amount of used memory and to
remove old knowledge that is no longer needed. Then, once the network acquired
the knowledge, the \emph{nextEnv} is stored as \emph{env} and a new \emph{mask}
is generated. Both \emph{env} and \emph{mask} are used to get an \emph{action}
according to the maximum value of the output the evaluation network. Once, the
action is determined, the \emph{nextEnv} is recreated for the next iteration.
The next iteration will occur, before the next service execution occurs, when
the model is asked again for a new decision.

\begin{figure*}
  \centering
  \begin{subfigure}[b]{0.32\textwidth}
    \includegraphics[width=\textwidth]{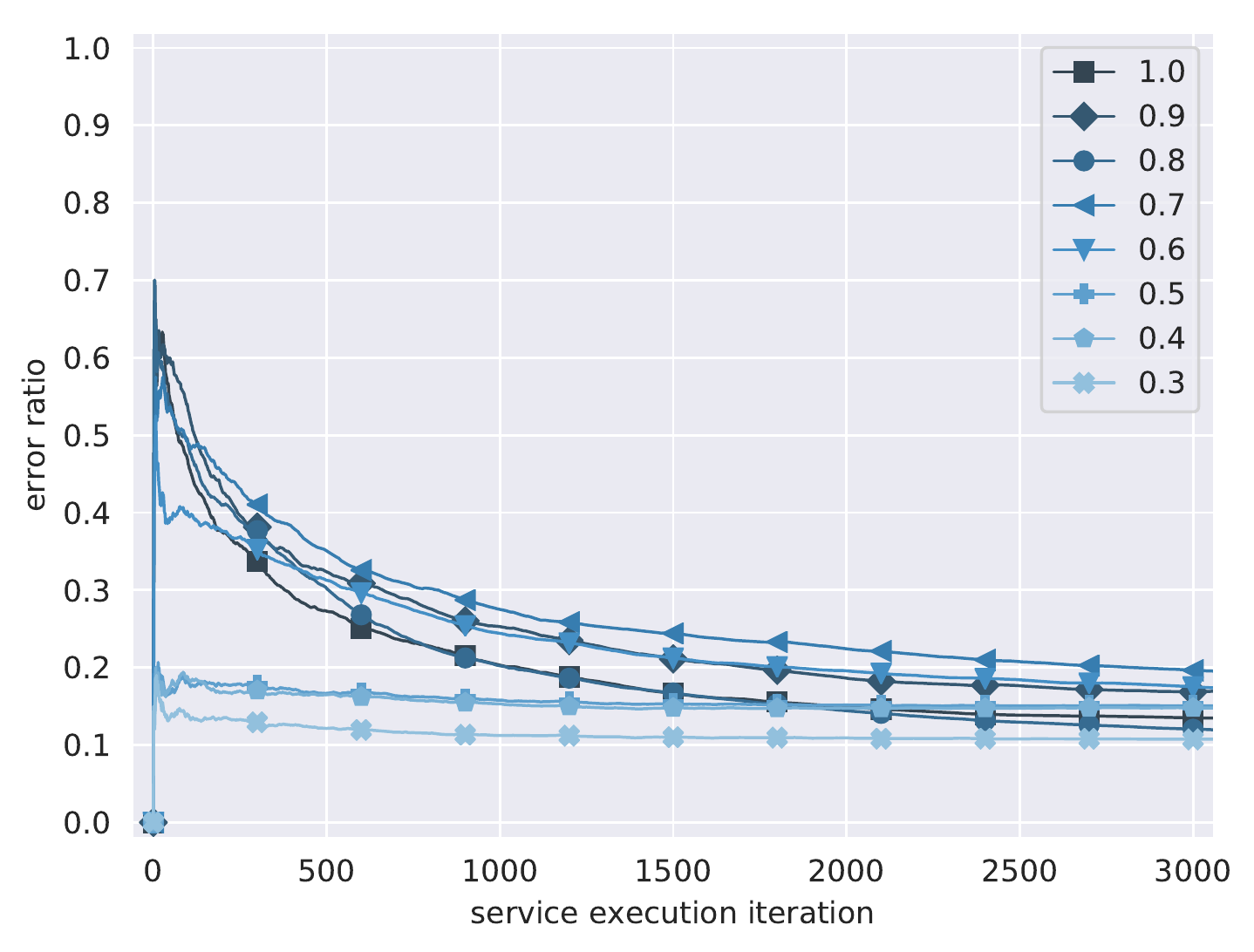}
    \caption{DRL - 5 devices}
    \label{fig:5_drl}
  \end{subfigure}
  \begin{subfigure}[b]{0.32\textwidth}
    \includegraphics[width=\textwidth]{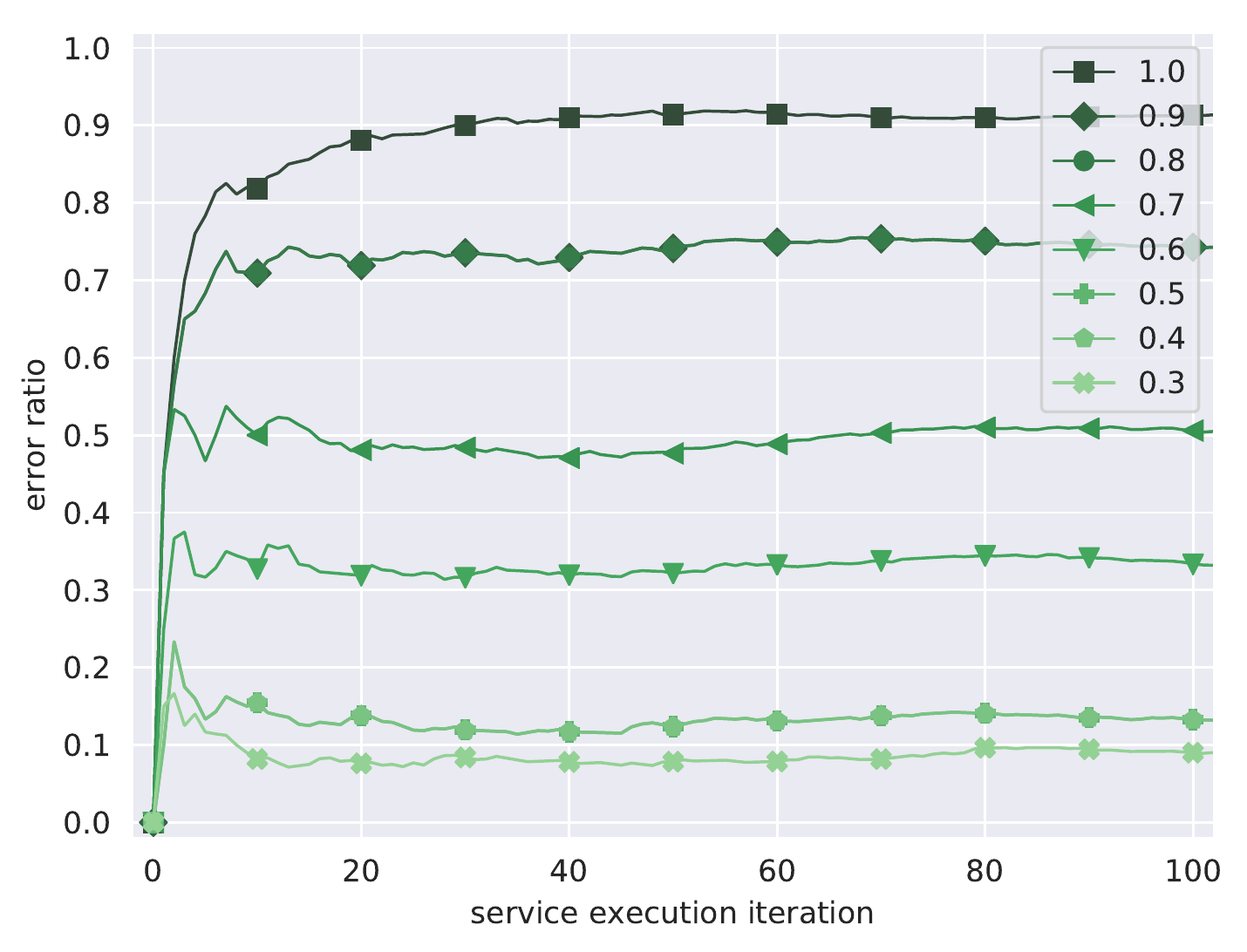}
    \caption{TEL - 5 devices}
    \label{fig:5_heu}
  \end{subfigure}
  \begin{subfigure}[b]{0.32\textwidth}
      \includegraphics[width=\textwidth]{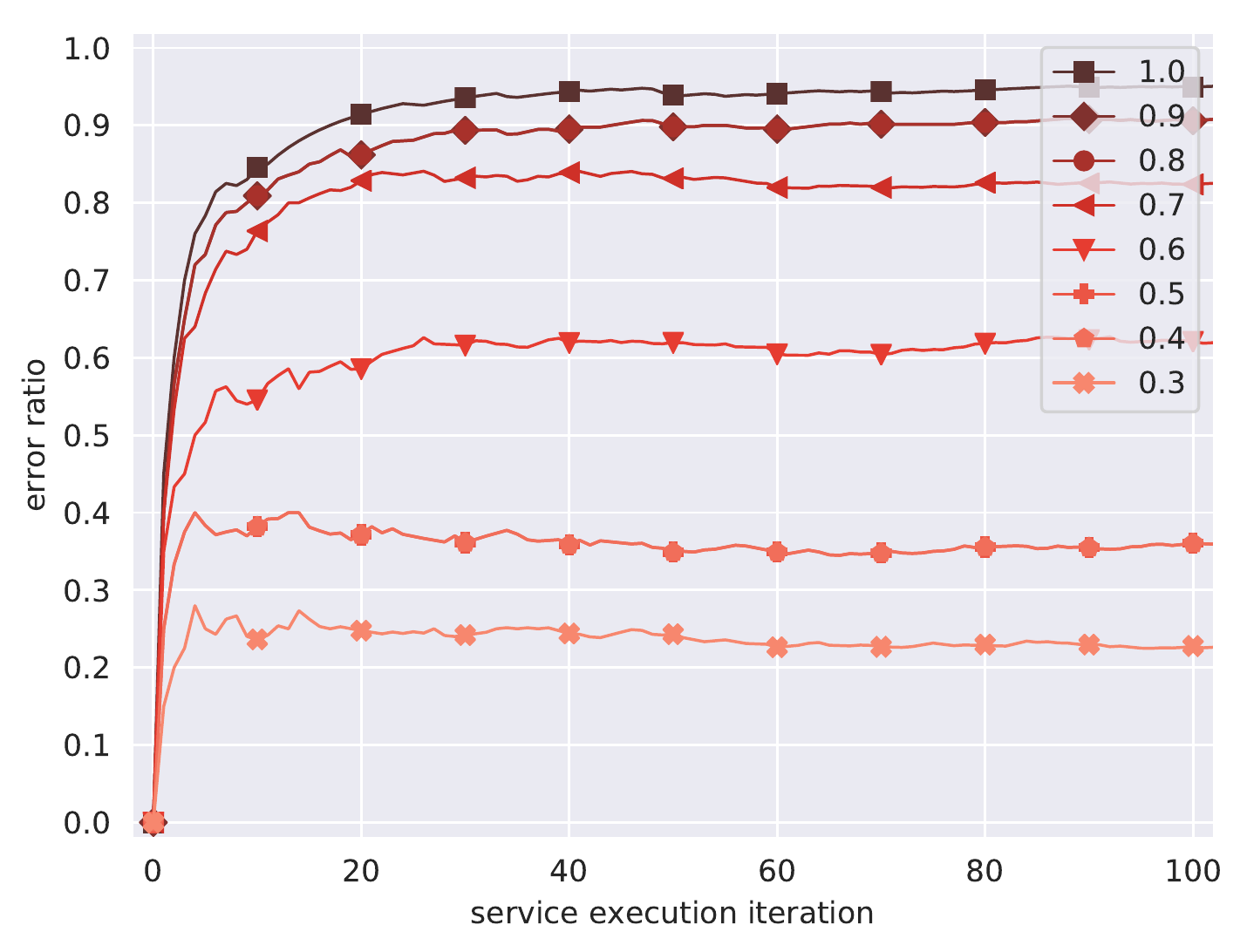}
      \caption{RND - 5 devices}
      \label{fig:5_rnd}
  \end{subfigure}
  \begin{subfigure}[b]{0.32\textwidth}
    \includegraphics[width=\textwidth]{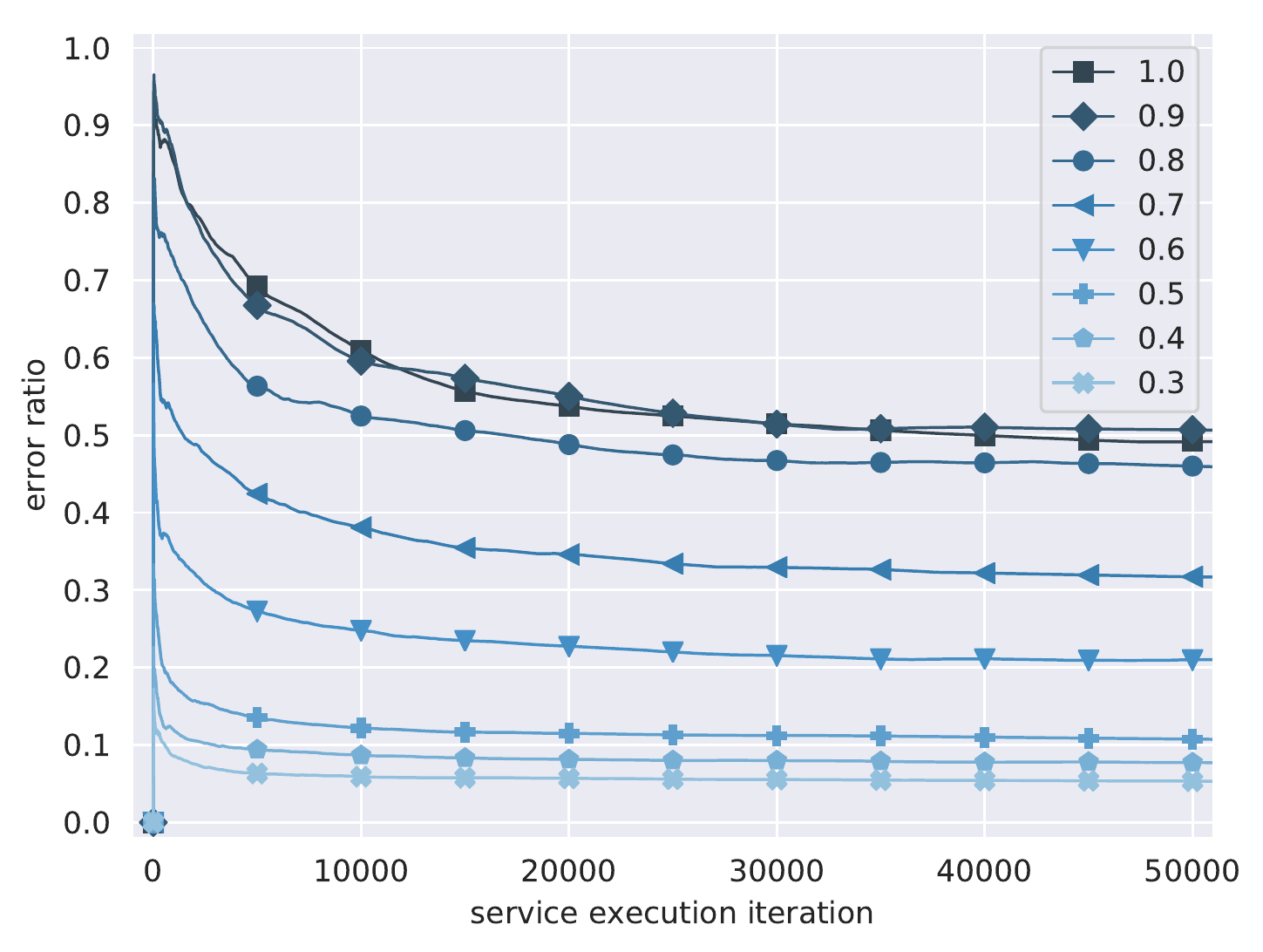}
    \caption{DRL - 10 devices}
    \label{fig:10_drl}
  \end{subfigure}
  \begin{subfigure}[b]{0.32\textwidth}
    \includegraphics[width=\textwidth]{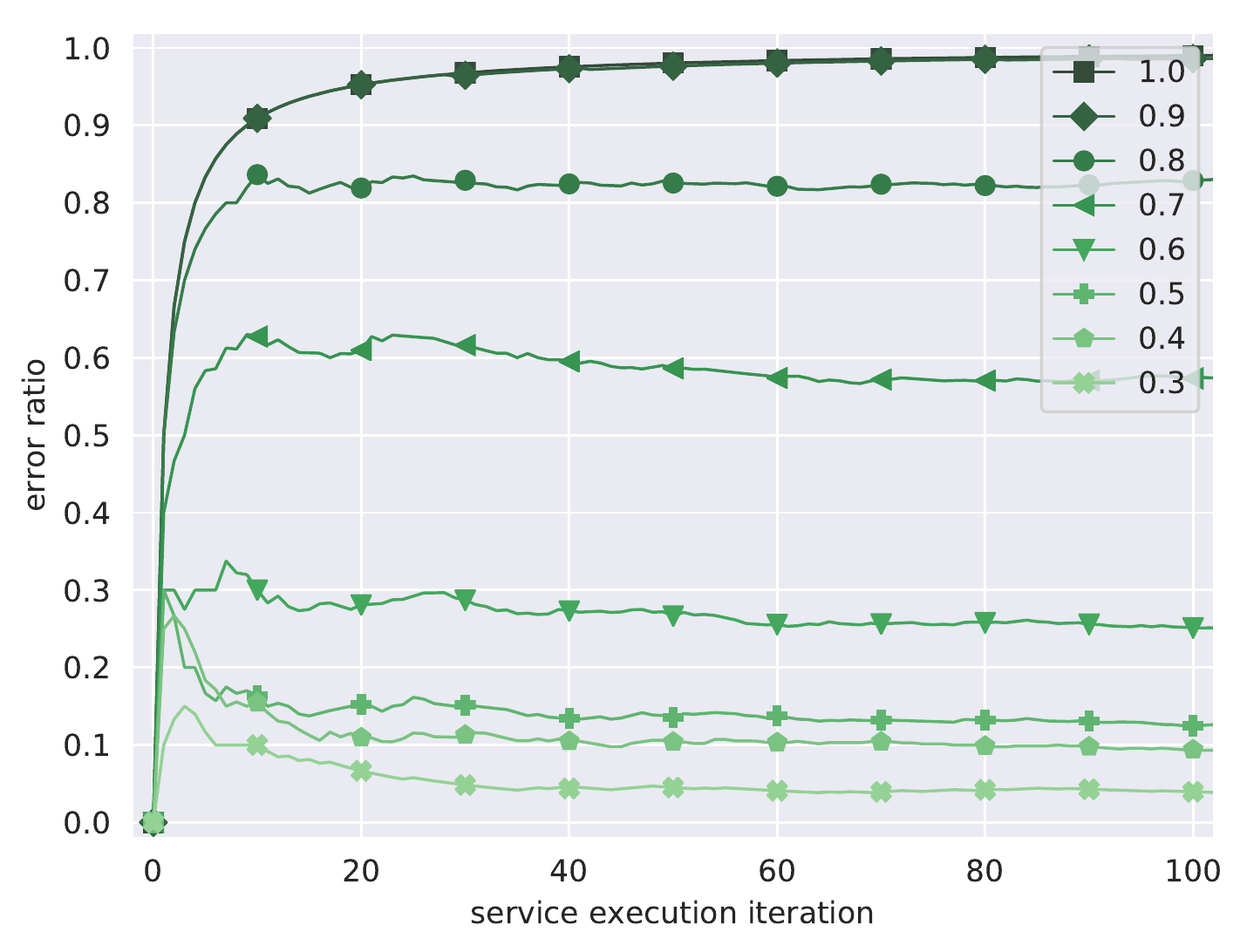}
    \caption{TEL - 10 devices}
    \label{fig:10_heu}
  \end{subfigure}
  \begin{subfigure}[b]{0.32\textwidth}
      \includegraphics[width=\textwidth]{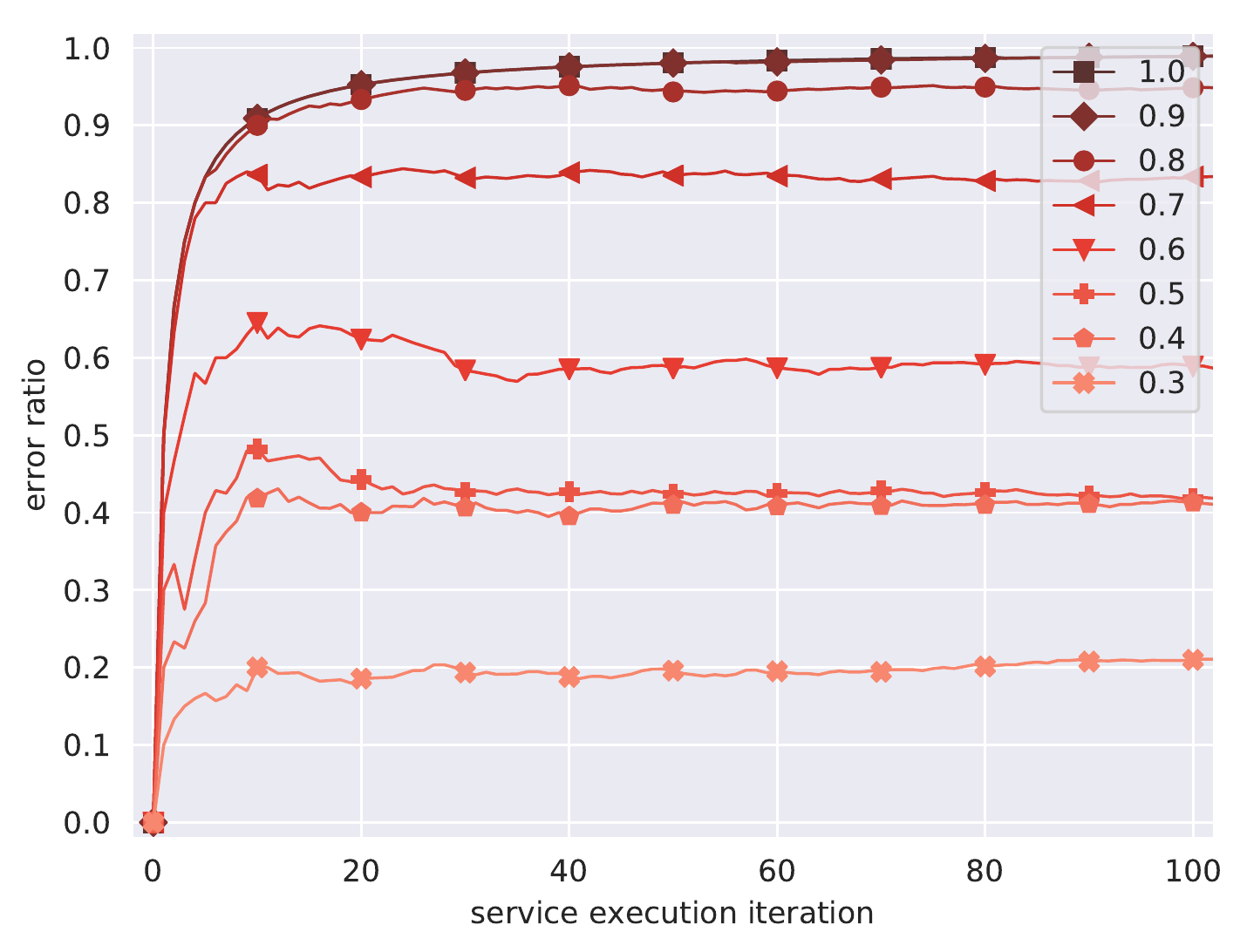}
      \caption{RND - 10 devices}
      \label{fig:10_rnd} 
  \end{subfigure}
  \begin{subfigure}[b]{0.32\textwidth}
    \includegraphics[width=\textwidth]{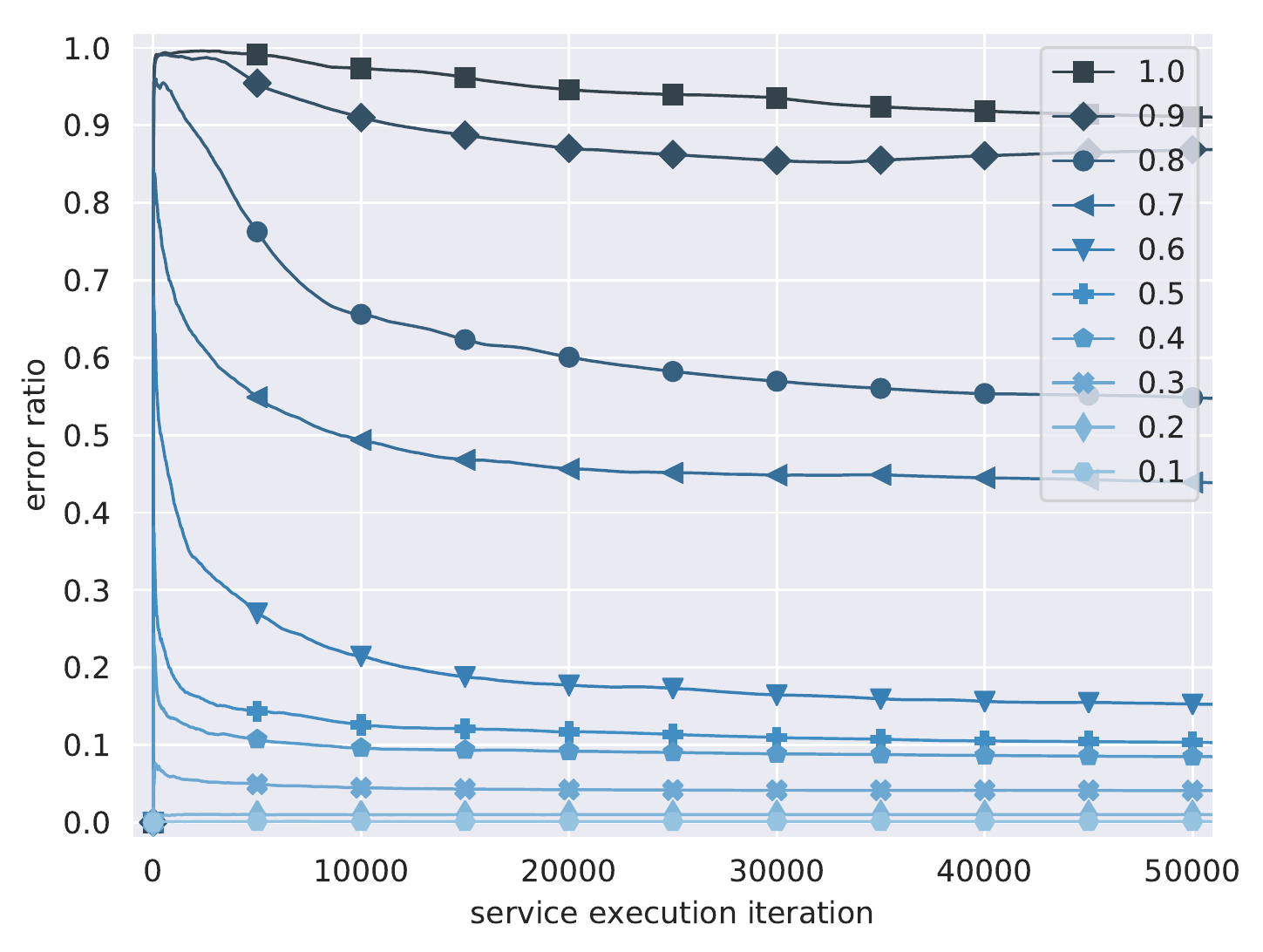}
    \caption{DRL - 15 devices}
    \label{fig:15_drl}
  \end{subfigure}
  \begin{subfigure}[b]{0.32\textwidth}
    \includegraphics[width=\textwidth]{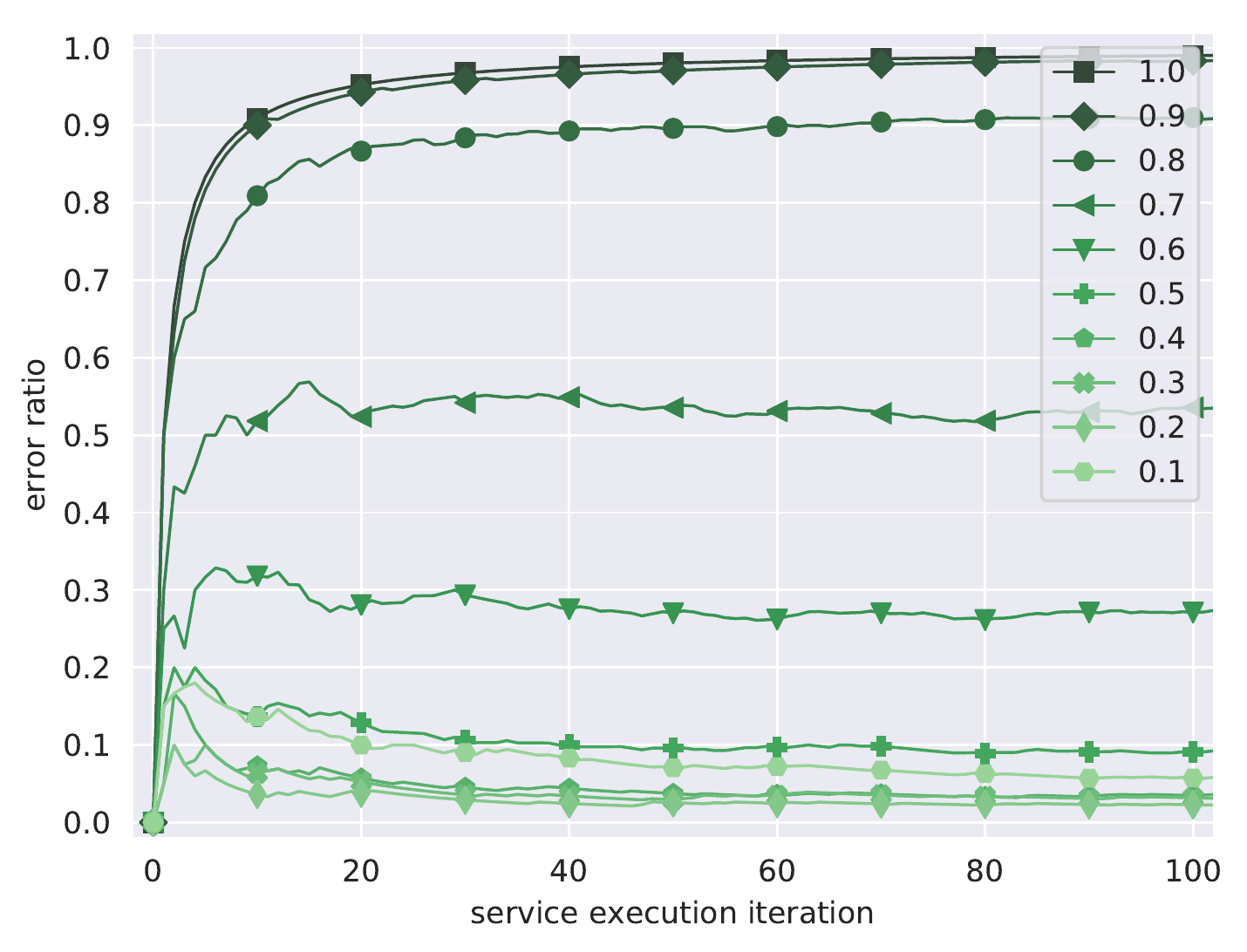}
    \caption{TEL - 15 devices}
    \label{fig:15_heu}
  \end{subfigure}
  \begin{subfigure}[b]{0.32\textwidth}
      \includegraphics[width=\textwidth]{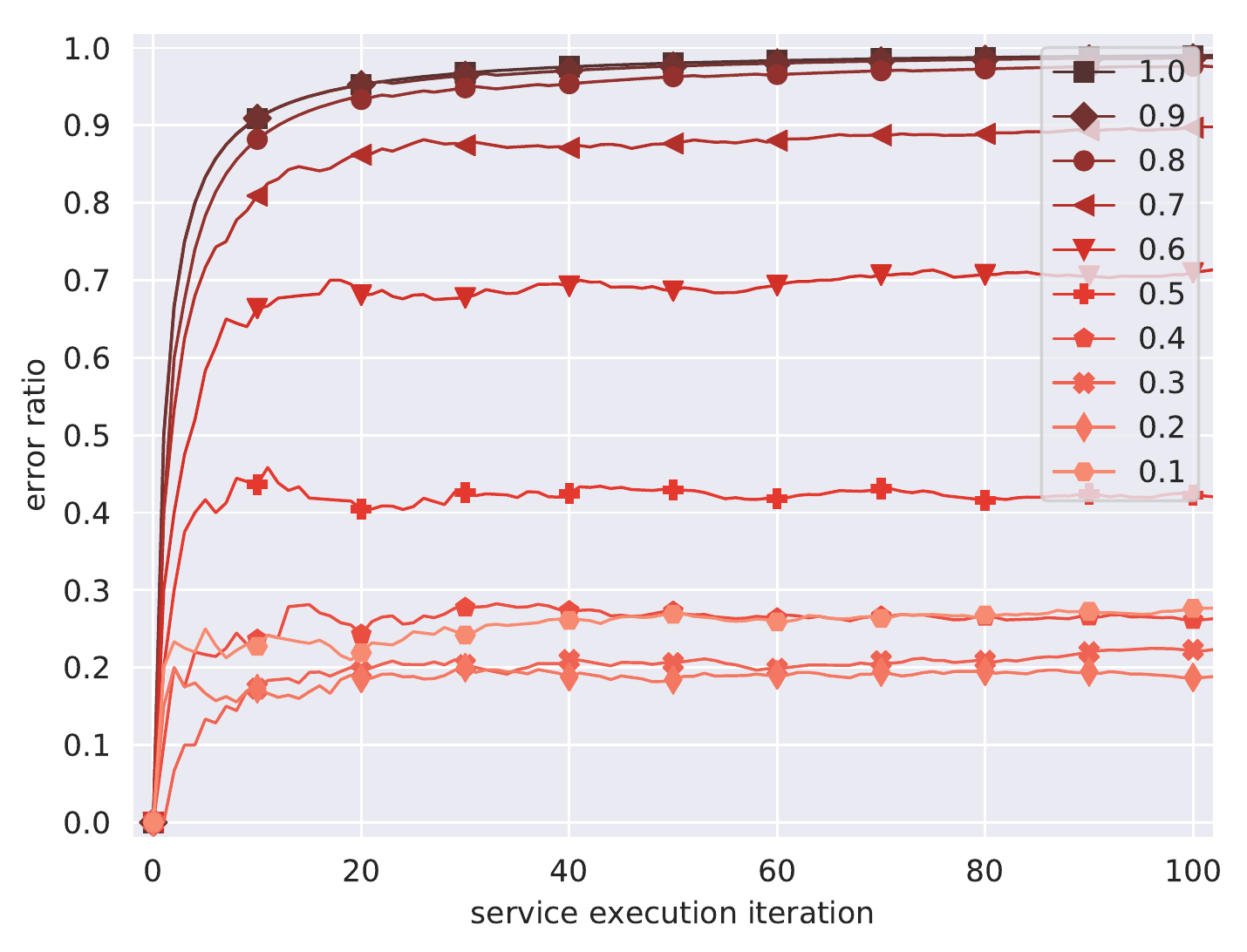}
      \caption{RND - 15 devices} 
      \label{fig:15_rnd}
  \end{subfigure}
  \caption{Error ratio running average per service executions}\label{fig:error_ratio}
  \vspace{-0.2cm}
\end{figure*}

\section{Performance Evaluation}

The software development described in this paper is open source and publicly
available in \cite{mf2c}. For the implementation of the \emph{DRL} algorithm we
have used Eclipse Deeplearning4j library, with the next parameters for the
model: \emph{memory\_capacity} is 100000, \emph{batch\_size} and
\emph{start\_size} are both 10, the \emph{discount\_factor} is 0.1 and the
\emph{num\_hidden\_layers} is 150. All tests have been performed in a Intel Core
i7-6700 CPU at 3.4GHz with 32 GB of RAM. For the sake of comparison, we also
implemented a random selection (RND) algorithm based, where the devices that are
block are randomly chosen. For all evaluations, instead of executing real
services, we just emulate the volatility of devices following a uniform
distributed function of the probability that the device fails, with the same
initial seeds for all three algorithms. To determine whether a service has been
disrupted or not, we use the \emph{acceptance\_ratio}. Therefore a service is
disrupted when the ratio of volatile devices divided by the total number of
devices used for the service execution is bigger than the acceptance ratio. For
instance, let us consider a cluster of 5 devices with an \emph{acceptance\_ratio
= 0.5} and all 5 devices are used for launching a service. If during the service
execution 3 or more devices are volatile, then the service will be disrupted.   

Fig. \ref{fig:error_ratio} shows the disruption ratio running average per
 service execution when clustering 5, 10 and 15 devices, and Table
 \ref{tab:avg_error} shows the total average service disruption ratio for 100k
 service executions. We evaluated the algorithms in all cases for acceptance
 ratios from 0.3 to 1.0. For the DRL algorithm, we show the running average for
 3000 service executions when clustering 5 devices and 50k service executions
 when clustering 10 and 15 devices. For the TEL and RND cases, we only show the
 first 100 executions, then the values become constant, however, the total
 average can be found in Table \ref{tab:avg_error}. When comparing all three
 algorithms when clustering 5 devices (see Fig \ref{fig:5_drl}, \ref{fig:5_heu}
 and \ref{fig:5_rnd}), we can see how DRL performs much better than TEL or RND
 for any acceptance ratio, even during the first service executions. The reason
 is related to the number of possible actions that DRL can take. Because DRL can
 only perform one action per service execution, when the number of possible
 actions (proportional to the number of devices) is low, the algorithm has more
 chances to predict the device that will fail. Instead, although TEL can block
 multiple devices per service execution, this blocking is only based on
 probability of failure. In RND case, the results are even worse than in TEL,
 because the blocking decision is randomly taken. When running the algorithms
 for 10 devices (see Fig. \ref{fig:10_drl}, \ref{fig:10_heu} and
 \ref{fig:10_rnd}) DRL still performs better in long term compared to the other
 two, but here we can see how this difference is less significant or negligible
 when the acceptance ratio is lower than 0.6 when compared to TEL (also in Table
 \ref{tab:avg_error}). This is because, the lower acceptance ratio the lower
 number of devices need to blocked, and then reducing the probability for the
 TEL to miss a device that will potential fail. The last case, when comparing
 the results for 15 devices, we can see how with acceptance ratio of 0.7 or
 lower, there is no benefit of using DRL over TEL, and only in long term (with
 more than 10k executions) in some cases DRL overperforms TEL. These results
 show how DRL solution performs much better with a low number of devices, due to
 the lower amount for actions from where the algorithm has to choose. For a high
 number of devices there is no benefit of using DRL instead of traditional TEL
 algorithms. We finally measured the average execution times over 100
 repetitions after 10 warmups for all algorithms. With 5 devices in the cluster
 DRL takes $333.797 \pm 2.965$ ms, TEL $0.153 \pm 0.010$ ms and RND $0.150 \pm
 0.001$ ms. For 10 devices DRL takes $336.241 \pm 34.058$ ms, TEL $0.328 \pm
 0.009$ ms and RND $0.312 \pm 0.010$ ms. For 15 devices DRL takes $394.507 \pm
 24.173$ ms, TEL $0.834 \pm 0.045$ ms and RND $0.523 \pm 0.029$ ms. We can see
 that DRL is much slower compared to TEL, but the amount of time is still
 negligible.

 \begin{table}[htbp]
  \centering
  \scriptsize
  \caption{Total average error ratio for 100k service executions}\label{tab:avg_error}
  \begin{tabular}{l|ccc|ccc|ccc}
  \cline{2-10}
  \multicolumn{1}{c|}{} & \multicolumn{3}{c|}{\textbf{5 devices}}    & \multicolumn{3}{c|}{\textbf{10 devices}}   & \multicolumn{3}{c}{\textbf{15 devices}}    \\ \cline{2-10} 
  \textbf{}             & DRL & TEL & RND & DRL & TEL & RND & DRL & TEL & RND \\ \hline
  1.0          & 0.08         & 0.92         & 0.96         & 0.47         & $0.\overline{99}$         & $0.\overline{99}$         & 0.90         & $0.\overline{99}$         & $0.\overline{99}$         \\ \hline
  0.9          & 0.11         & 0.75         & 0.91         & 0.52         & 0.99         & $0.\overline{99}$         & 0.87         & 0.99         & $0.\overline{99}$         \\ \hline
  0.8          & 0.08         & 0.75         & 0.91         & 0.45         & 0.84         & 0.95         & 0.53         & 0.90         & 0.99         \\ \hline
  0.7          & 0.15         & 0.52         & 0.82         & 0.31         & 0.57         & 0.85         & 0.41         & 0.51         & 0.89         \\ \hline
  0.6          & 0.10         & 0.33         & 0.63         & 0.20         & 0.23         & 0.60         & 0.13         & 0.22         & 0.72         \\ \hline
  0.5          & 0.15         & 0.12         & 0.38         & 0.10         & 0.10         & 0.43         & 0.10         & 0.05         & 0.43         \\ \hline
  0.4          & 0.13         & 0.12         & 0.38         & 0.07         & 0.08         & 0.43         & 0.08         & 0.02         & 0.27         \\ \hline
  0.3          & 0.10         & 0.08         & 0.25         & 0.05         & 0.02         & 0.20         & 0.04         & 0.01         & 0.22         \\ \hline
  \end{tabular}
  \vspace{-0.3cm}
  \end{table}

\section{Conclusion}

We proposed a QoS provider mechanism, as an integral component of a real world
fog-to-cloud system, to work in dynamic edge computing scenarios based on
reinforcement learning. Specifically, we developed a deep Q-learning algorithm
which is particularly well suited in dynamic and adaptive environments where the
decision process needs to be frequently updated. We compared our solution with a
telemetry based heuristic algorithm, showing how reinforcement learning is able
to overperform when the number of devices to manage is low. As future work, we
will extend our algorithm to allow multiple actions per service execution,
expecting to improve the results when increasing the number of managed devices.

\bibliographystyle{IEEEtran}
\bibliography{qos-paper}

\begin{thebibliography}{10}
\providecommand{\url}[1]{#1}
\csname url@samestyle\endcsname
\providecommand{\newblock}{\relax}
\providecommand{\bibinfo}[2]{#2}
\providecommand{\BIBentrySTDinterwordspacing}{\spaceskip=0pt\relax}
\providecommand{\BIBentryALTinterwordstretchfactor}{4}
\providecommand{\BIBentryALTinterwordspacing}{\spaceskip=\fontdimen2\font plus
\BIBentryALTinterwordstretchfactor\fontdimen3\font minus
  \fontdimen4\font\relax}
\providecommand{\BIBforeignlanguage}[2]{{%
\expandafter\ifx\csname l@#1\endcsname\relax
\typeout{** WARNING: IEEEtran.bst: No hyphenation pattern has been}%
\typeout{** loaded for the language `#1'. Using the pattern for}%
\typeout{** the default language instead.}%
\else
\language=\csname l@#1\endcsname
\fi
#2}}
\providecommand{\BIBdecl}{\relax}
\BIBdecl

\bibitem{Faniyi2015}
F.~Faniyi and R.~Bahsoon, ``{A Systematic Review of Service Level Management in
  the Cloud},'' \emph{ACM Computing Surveys}, 2015.

\bibitem{mf2c}
\BIBentryALTinterwordspacing
H2020, ``{mF2C: Towards an Open, Secure, Decentralized and Coordinated
  Fog-to-Cloud Management Ecosystem}.'' [Online]. Available:
  \url{http://www.mf2c-project.eu}
\BIBentrySTDinterwordspacing

\bibitem{openfog}
\BIBentryALTinterwordspacing
O.~Consortium, ``{OpenFog Consortium},'' 2017. [Online]. Available:
  \url{https://www.openfogconsortium.org/}
\BIBentrySTDinterwordspacing

\bibitem{Kan2018}
T.~Y. Kan, Y.~Chiang, and H.~Y. Wei, ``{Task offloading and resource allocation
  in mobile-edge computing system},'' \emph{2018 27th Wireless and Optical
  Communication Conference, WOCC 2018}, pp. 1--4, 2018.

\bibitem{Song2017}
Y.~Song, S.~S. Yau, R.~Yu, X.~Zhang, and G.~Xue, ``{An Approach to QoS-based
  Task Distribution in Edge Computing Networks for IoT Applications},''
  \emph{Proceedings - 2017 IEEE 1st International Conference on Edge Computing,
  EDGE 2017}, pp. 32--39, 2017.

\bibitem{Huang2017}
L.~Huang, G.~Li, J.~Wu, L.~Li, J.~Li, and R.~Morello, ``{Software-defined QoS
  provisioning for fog computing advanced wireless sensor networks},''
  \emph{Proceedings of IEEE Sensors}, pp. 1--3, 2017.

\bibitem{Li2010}
D.~H. Li and D.~Levy, ``{A reinforcement learning based self-optimizing QoS
  controller framework for distributed services},'' \emph{2010 Chinese Control
  and Decision Conference, CCDC 2010}, pp. 2917--2922, 2010.

\bibitem{Khelifi2019}
H.~Khelifi, S.~Luo, B.~Nour, A.~Sellami, H.~Moungla, S.~H. Ahmed, and
  M.~Guizani, ``{Bringing Deep Learning at the Edge of Information-Centric
  Internet of Things},'' \emph{IEEE Communications Letters}, 2019.

\bibitem{Wang2019}
J.~Wang, L.~Zhao, J.~Liu, and N.~Kato, ``{Smart Resource Allocation for Mobile
  Edge Computing: A Deep Reinforcement Learning Approach},'' \emph{IEEE
  Transactions on Emerging Topics in Computing}, vol.~PP, no.~c, pp. 1--1,
  2019.

\bibitem{Abundo2015}
M.~Abundo, V.~{Di Valerio La Sapienza}, V.~Cardellini, and F.~L. Presti,
  ``{QoS-aware bidding strategies for VM spot instances: A reinforcement
  learning approach applied to periodic long running jobs},'' \emph{Proceedings
  of the 2015 IFIP/IEEE International Symposium on Integrated Network
  Management, IM 2015}, pp. 53--61, 2015.

\bibitem{Lin2016}
S.~C. Lin, I.~F. Akyildiz, P.~Wang, and M.~Luo, ``{QoS-aware adaptive routing
  in multi-layer hierarchical software defined networks: A reinforcement
  learning approach},'' \emph{Proceedings - 2016 IEEE International Conference
  on Services Computing, SCC 2016}, pp. 25--33, 2016.

\bibitem{Wei2018}
Y.~Wei, L.~Pan, S.~Liu, L.~Wu, and X.~Meng, ``{DRL-Scheduling: An intelligent
  QoS-Aware job scheduling framework for applications in clouds},'' \emph{IEEE
  Access}, vol.~6, pp. 55\,112--55\,125, 2018.

\bibitem{Andrieux2004}
\BIBentryALTinterwordspacing
A.~Andrieux, K.~Czajkowski, A.~Dan, and K.~Keahey, ``{Web services agreement
  specification (WS-Agreement)},'' Tech. Rep. [Online]. Available:
  \url{https://www.ggf.org/Public{\_}Comment{\_}Docs/Documents/Oct-2006/WS-AgreementSpecificationDraftFinal{\_}sp{\_}tn{\_}jpver{\_}v2.pdf}
\BIBentrySTDinterwordspacing

\end{thebibliography}

\end{document}